\documentclass[sigconf]{acmart}

\AtBeginDocument{%
  \providecommand\BibTeX{{%
    \normalfont B\kern-0.5em{\scshape i\kern-0.25em b}\kern-0.8em\TeX}}}

\copyrightyear{2022}
\acmYear{2022}
\setcopyright{acmcopyright}\acmConference[WWW '22]{Proceedings of the ACM Web Conference 2022}{April 25--29, 2022}{Virtual Event, Lyon, France}
\acmBooktitle{Proceedings of the ACM Web Conference 2022 (WWW '22), April 25--29, 2022, Virtual Event, Lyon, France}
\acmPrice{15.00}
\acmDOI{10.1145/3485447.3512258}
\acmISBN{978-1-4503-9096-5/22/04}

\acmSubmissionID{w12fp2892}

\usepackage{graphicx}
\usepackage{caption}
\usepackage{subcaption}
\usepackage{multirow}
\usepackage{algorithm}
\usepackage{amsmath}
\usepackage{algorithmic}

\usepackage{float}
\usepackage[normalem]{ulem}
\usepackage{multirow}
\usepackage{booktabs}
\usepackage{xspace}

\newcommand{\m}{\textsc{REAL-FND}\xspace}
\begin{document}

\title{Domain Adaptive Fake News Detection via Reinforcement Learning}


\author{Ahmadreza Mosallanezhad}
\email{amosalla@asu.edu}
\affiliation{%
  \institution{Arizona State University}
  \city{Tempe}
  \state{AZ}
  \country{USA}
}

\author{Mansooreh Karami}
\email{mkarami@asu.edu}
\affiliation{%
  \institution{Arizona State University}
  \city{Tempe}
  \state{AZ}
  \country{USA}
}

\author{Kai Shu}
\email{kshu@iit.edu}
\affiliation{%
  \institution{Illinois Institute of Technology}
  \city{Chicago}
  \state{IL}
  \country{USA}
}

\author{Michelle V. Mancenido}
\email{mmanceni@asu.edu}
\affiliation{%
  \institution{Arizona State University}
  \city{Glendale}
  \state{AZ}
  \country{USA}
}

\author{Huan Liu}
\email{huanliu@asu.edu}
\affiliation{%
  \institution{Arizona State University}
  \city{Tempe}
  \state{AZ}
  \country{USA}
}

\renewcommand{\shortauthors}{Mosallanezhad et al.}

\begin{abstract}

With social media being a major force in information consumption, accelerated propagation of fake news has presented new challenges for platforms to distinguish between legitimate and fake news. Effective fake news detection is a non-trivial task due to the diverse nature of news domains and expensive annotation costs. In this work, we address the limitations of existing automated fake news detection models by  incorporating auxiliary information (e.g., user comments and user-news interactions) into a novel reinforcement learning-based model called \textbf{RE}inforced \textbf{A}daptive \textbf{L}earning \textbf{F}ake \textbf{N}ews \textbf{D}etection (REAL-FND). REAL-FND exploits cross-domain and within-domain knowledge that makes it robust in a target domain, despite being trained in a different source domain. Extensive experiments on real-world datasets illustrate the effectiveness of the proposed model, especially when limited labeled data is available in the target domain.

\end{abstract}


\maketitle

\section{Introduction}

With people spending more time on social media platforms\footnote{In 2020, internet users spent an average of 145 minutes per day on social media~\cite{statista2021}.}, it is not surprising that social media has evolved into the primary source of news among subscribers, in lieu of the more traditional news delivery systems such as early morning shows, newspapers, and websites affiliated with media companies. For example, it was reported that 1 in 5 U.S. adults used social media as their primary source of political news during the 2020 U.S. presidential elections~\cite{amymitchell2020}. 
\begin{figure}[t]
\centering
\begin{subfigure}{.25\textwidth}
  \centering
  \includegraphics[width=0.8\linewidth]{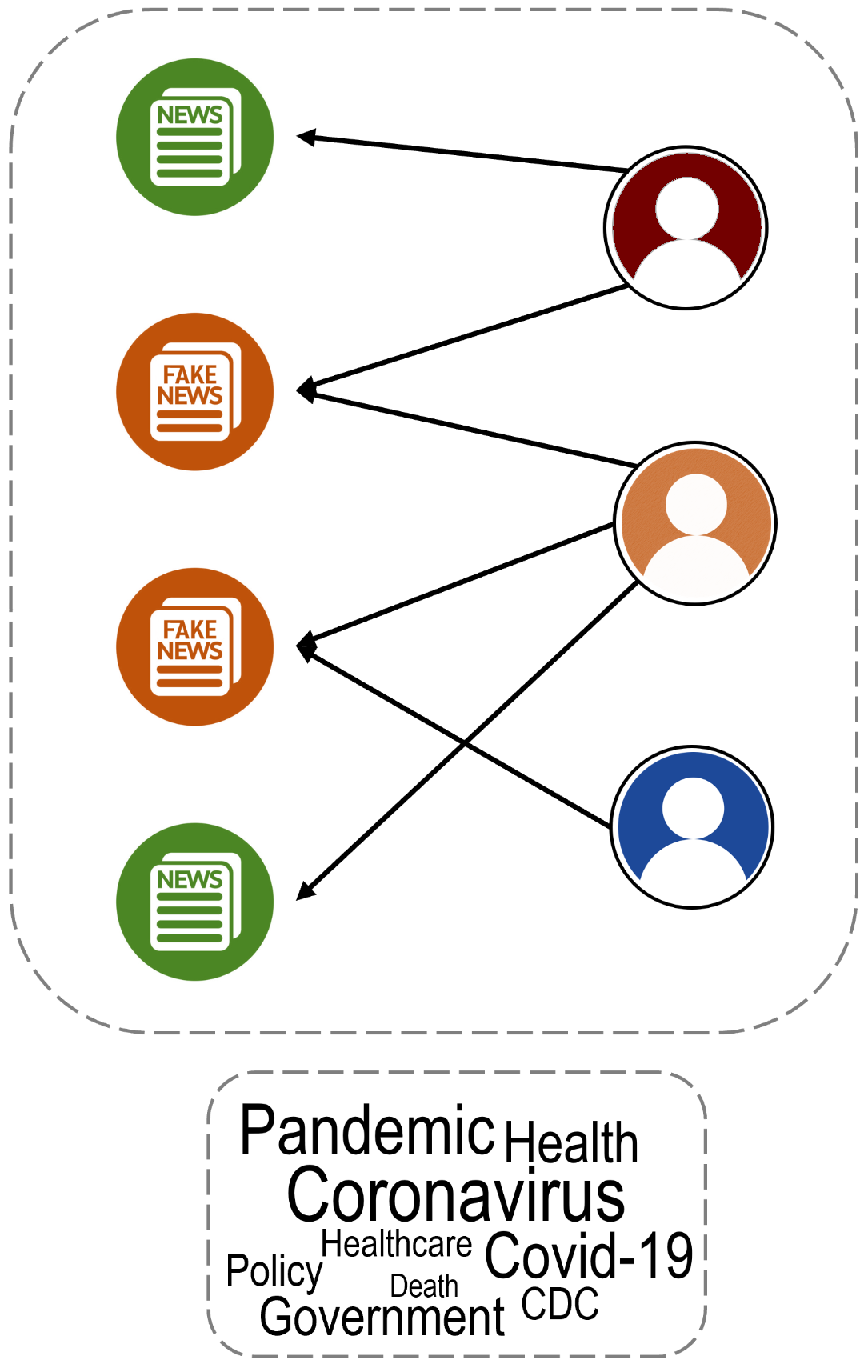}
  \caption{Source Domain - Healthcare}
  \label{fig:sub1}
\end{subfigure}%
\begin{subfigure}{.25\textwidth}
  \centering
  \includegraphics[width=0.8\linewidth]{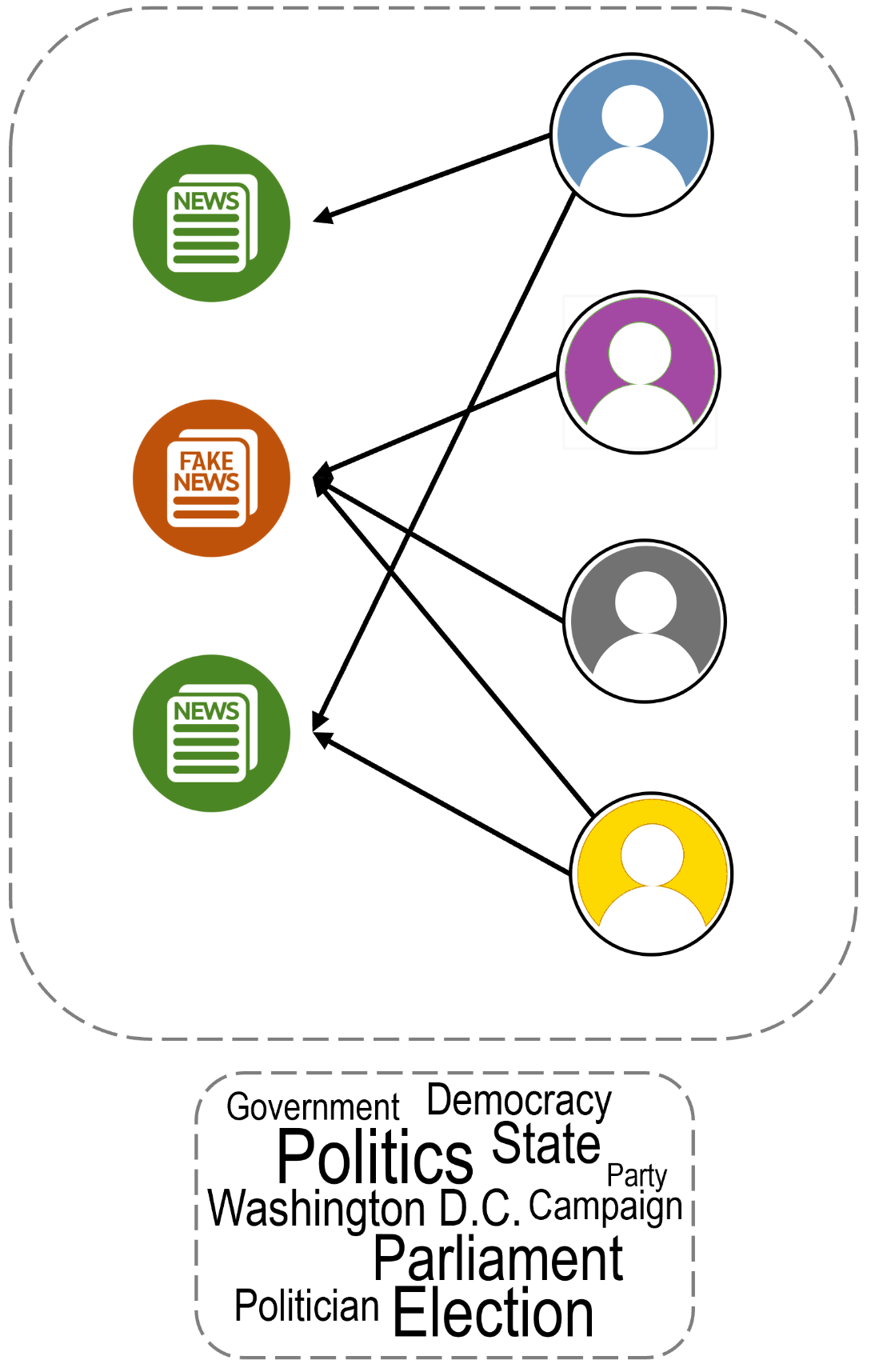}
  \caption{Target Domain - Politics}
  \label{fig:sub2}
\end{subfigure}
\caption{Comparing two domains, Healthcare and Politics, in fake news detection: (1)~Different word usage, (2)~Event-related words/features, (3)~Different user-news interaction. }
\label{fig:DsDt}
\end{figure}
The ease and speed of disseminating new information via social media, however, have created networks of disinformation that propagate as fast as any social media post. For example from the COVID pandemic era, around 800 fatalities, 5000 hospitalizations, and 60 permanent injuries were recorded as a result of false claims that household bleach was an effective panacea for the SARS-CoV-2 virus~\cite{islam2020covid, coleman_2020}. Unlike traditional news delivery where trained reporters and editors fact-check information, curation of news on social media has largely been crowd-sourced, i.e., social media users themselves are the producers and consumers of information. 

In general, humans have been found to fare worse than machines at prediction tasks~\cite{10.1162/coli_a_00380,shabani2018hybrid} such as distinguishing between fake or legitimate news. Machine learning models for automated fake news detection have even been shown to perform better than the most seasoned linguists~\cite{perez-rosas-etal-2018-automatic}. To this end, many automated fake news detection algorithms have largely focused on improving predictive performance for a specific news domain (e.g., political news). The primary issue with these existing state-of-the-art detection algorithms is that while they perform well for the domain they were trained on (e.g., politics), they perform poorly in other domains (e.g., healthcare). 
The limited cross-domain effectiveness of algorithms to detect fake news are mostly due to (1)~the reliance of content-based approaches on word usage that are specific to a domain, (2)~the model's bias towards event-specific features, and (3)~domain-specific user-news interaction patterns (Figure~\ref{fig:DsDt}). As one of the contributions of this work, we will empirically demonstrate that the advertised performance of SOTA methods is not robust across domains.

Additionally, due to the high cost and specialized expertise required for data annotation, limited training data is available for effectively training an automated model across domains. This calls for using auxiliary information such as users' comments and motivational factors~\cite{shu2019beyond, karami2021profiling} as value-adding pieces for fake news detection.  

To address these challenges, we propose a domain-adaptive model called \textbf{RE}inforced \textbf{A}daptive \textbf{L}earning \textbf{F}ake \textbf{N}ews \textbf{D}etection (REAL-FND), which uses generalized and domain-independent features to distinguish between fake and legitimate news. The proposed model is based on previous evidence that illustrates how domain-invariant features could be used to improve the robustness and generality of fake news detection methods. As an example of a domain-invariant feature, it has been shown that fake news publishers use click-bait writing styles to attract specific audiences~\cite{zhou2020fake}. On the other hand,  patterns extracted from the social context provide rich information for fake news classification within a domain. For example, a user's comment providing evidence in refuting a piece of news is a valuable source of auxiliary information~\cite{shu2019defend}. Or, if a specific user is a tagged fake news propagator, the related user-news interaction could be leveraged as an additional source of information~\cite{karami2021profiling}.

In REAL-FND, instead of applying the commonly-used method of adversarial learning in training the cross-domain model, we transform the learned representation from the source to the target domain by deploying a reinforcement learning (RL) component. Other RL-based methods employ the agent to modify the parameters of the model. However, we use the RL agent to modify the learned representations to ensure that domain-specific features are obscured while domain-invariant components are maintained. An RL agent provides more flexibility over adversarial training because any classifier's confidence values could be directly optimized (i.e., the objective function does not need to be differentiable).

We address the challenges in domain adaptive fake news detection algorithms by making the following contributions:
\begin{itemize}
    \item We design a framework that encodes news content, user comments, and user-news interactions as representation vectors and fuses these representations for fake news detection;
    \item We utilize a reinforcement learning setting to adjust the representations to account for domain-independent features;
    \item We conduct extensive experiments on real-world datasets to show the effectiveness of the proposed model, especially when limited labeled data is available in the target domain.
\end{itemize}

\section{Related Work}
The proposed methodology spans the subject domains of fake news detection, cross domain modeling, and the use of reinforcement learning in the Natural Language Processing (NLP). Current state-of-the-art in these areas are discussed in this section.  

\subsection{Fake News Detection}
Early research in fake news detection focused on extracting features from news content to either capture the textual information (such as lexical and syntactic features) ~\cite{shu2017fake, shu2020combating} or to detect the differences in the writing style of real and fake news such as deception~\cite{feng2012syntactic} and non-objectivity~\cite{potthast2017stylometric}. Qian et al. propose to use convolutional neural network to detect fake news based on news content only~\cite{qian2018neural}. Guo et al. propose to capture the sensational emotions to create an emotion-enhanced news representation to detect fake news~\cite{guo2019exploiting}. Recent advancements utilize deep neural networks~\cite{karimi2019learning} or tensor factorization~\cite{hosseinimotlagh2018unsupervised} to model latent news representations.

While news content-based approaches result in acceptable performance, incorporating auxiliary information improves the performance and reliability of fake news detection models~\cite{shu2019defend}. For example, Tacchini et al.~\cite{tacchini2017some} and Guo et al.~\cite{guo2018rumor} aggregated users' social responses (topics, stances, and credibility), Ruchansky et al. proposed a model called CSI which uses a hybrid model on news content and users network to detect fake news~\cite{ruchansky2017csi}, and Shu et al. used hierarchical attention networks to create more explainable fake news detection models~\cite{shu2019defend}. In this work, we incorporate various auxiliary information (i.e., users' comments and user-news interactions) along with the news content to create a well-defined news article representation.   

\subsection{Cross Domain Modeling}
Cross-domain modeling refers to a model capable of learning information from data in the source domain and being able to transfer it to a target domain~\cite{ben2010theory}. In general, cross-domain models are categorized into \textit{sample-level} and \textit{feature-level} groups~\cite{zhuang2015supervised}. Sample-level domain adaptation methods focus on finding domain-independent samples by assigning weights to these instances ~\cite{zhuang2015supervised,silva2021embracing}. On the other hand, feature-level domain adaptation methods focus on weighting or extracting domain-independent features~\cite{kouw2016feature}. In addition to the aforementioned domain adaptation methods, Gong et al. combined both sample-level and feature-level domain adaptation~\cite{gong2020unified} in BERT to create a domain-independent sentiment analysis model. Similarly, Vlad et al. used transfer learning on an enhanced BERT architecture to detect propaganda across domains~\cite{vlad2019sentence}. Moreover, Zhuang et al. propose to use auto-encoders for learning unsupervised feature representations for domain adaptation~\cite{zhuang2015supervised}. The goal of this model is to leverage a small portion of the target domain data to train an auto-encoder for learning domain-independent feature representations. 
In this paper, we focus on feature-level domain adaptation by leveraging reinforcement learning to learn a domain-independent textual representation.

\subsection{Reinforcement Learning}
In past years, reinforcement learning has shown to be useful in improving the effectiveness of NLP models. As such, Li et al. propose a method for generating paraphrases using inverse reinforcement learning~\cite{li2017paraphrase}. Another work by Fedus et al. uses reinforcement learning to tune parameters of an LSTM-based generative adversarial network for text generation~\cite{fedus2018maskgan}. A recent work by Cheng et al. proposes to use reinforcement learning for tuning a classifier's parameters for removing various types of biases~\cite{cheng2021mitigating}.
Inspired by these methods, we study the problem of domain-adaptive fake news detection using reinforcement learning. 

In previous research, although reinforcement learning has been used to modify a model's parameters, little research has been done on applying it to modify the learned representations. In this work, 
we focus on designing a domain adaptation model with an RL agent that modifies the article's representation based on the feedback received from both the adversary domain classifier and the fake news detection component.

\section{Problem Statement}
Let $\mathcal{D}_{s}=\{$ $(\mathbf{x}_1^s, y_1^s)$, $(\mathbf{x}_2^s, y_2^s)$, $...$, $(\mathbf{x}_N^s, y_N^s) \}$ and $\mathcal{D}_{t}= \{$ $(\mathbf{x}_1^t, y_1^t)$, $(\mathbf{x}_2^t, y_2^t)$, $...$, $(\mathbf{x}_M^t, y_M^t) \}$ denote a set of $N$ and $M$ news article with labels from source and target domains, respectively. Each news article $\mathbf{x}_i$ includes a content which is a sequence of $K$ words $\{ w_1, w_2, ..., w_K \}$, a set of comments $\mathbf{c}_j \in \mathcal{C}$, and user-news interactions $\mathbf{u}_j \in \mathcal{U}$. User-news interactions $\mathbf{u}_i$ is a binary vector indicating users who posted, re-posted, or liked a tweet about news $\mathbf{x}_i$. The goal of the reinforced domain adaptive agent is to learn a function that converts the news representation $\mathbf{x}_i \in \mathcal{D}_{t}$ from target domain to source domain. The problem is formally defined as follows:

\begin{figure*}[ht!]
    \centering
    \includegraphics[width=0.95\linewidth]{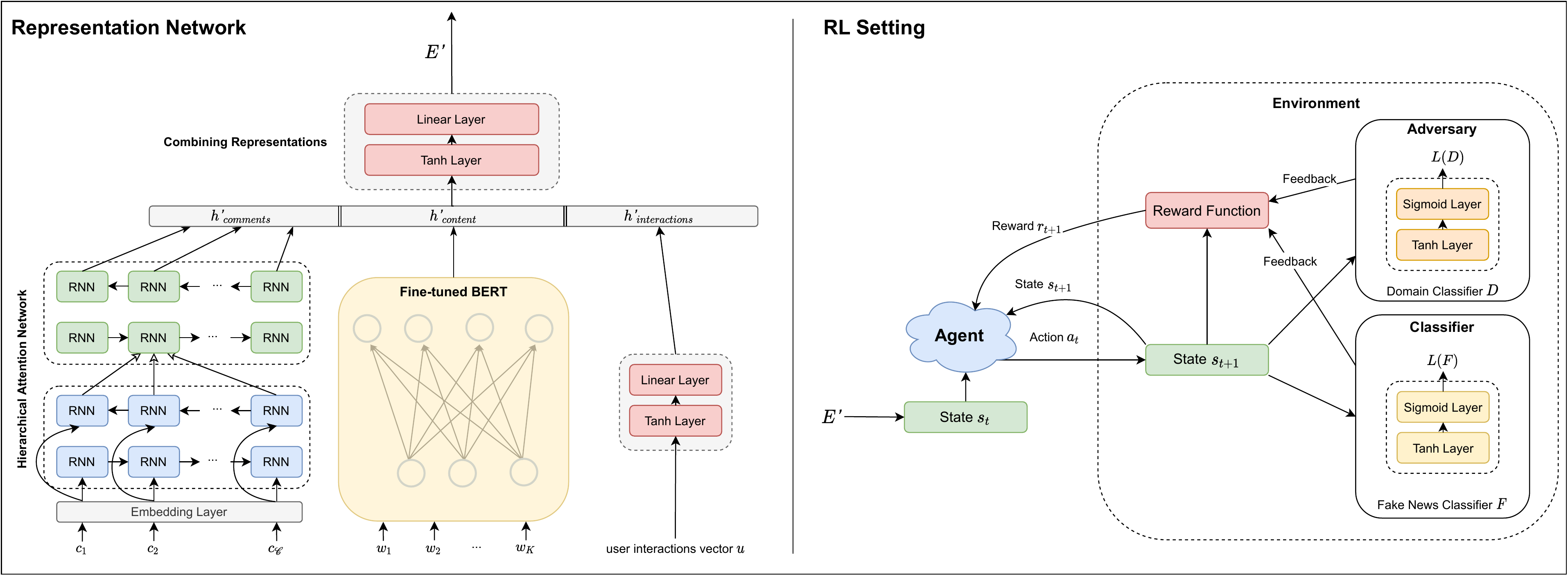}
    \caption{The proposed \m architecture. \m has two main components, the representation network, and the reinforcement learning agent. The representation network learns a representation including information about news content, comments, and user-news interactions. The RL agent learns a method to transfer the learned  representation from source domain to the target domain. }
    \label{fig:proposed}
\end{figure*}

\begin{center}
\fbox{\parbox[c]{.95\linewidth}{\textbf{Definition (Domain Adaptive Fake News Detection).} Given news articles from two separate domains $\mathcal{D}_s$ and $\mathcal{D}_t$, corresponding users' comments $\mathcal{C}_s$ and $\mathcal{C}_t$, and user-news interactions $\mathcal{U}_s$ and $\mathcal{U}_t$ from the source $S$ and target $T$ domains, respectively, learn a domain-independent news article representation using $\mathcal{D}_s$ and a small portion of $\mathcal{D}_t$ that can be classified correctly by the fake news classifier $F$.}}
\end{center}

\section{Proposed Model}

In this section, we describe our proposed model, \textbf{RE}inforced domain \textbf{A}daptation \textbf{L}earning for \textbf{F}ake \textbf{N}ews \textbf{D}etection~(\m). The input of this model is the news articles from source domain $\mathcal{D}_s$ and a portion ($\gamma$) of the target data set $\mathcal{D}_t$. As shown in \autoref{fig:proposed}, the \m model has two main components:~(1)~the news article encoder, and (2)~the reinforcement learning agent. In the following subsections, we explain these two components in detail.

\subsection{News Article Encoder}
The problem of detecting fake news requires learning a comprehensive representation that includes information about news content and its related auxiliary information. In this paper, we consider news comments and user-news interactions as auxiliary information for fake news detection. Recurrent neural networks (RNN) such as Long Short Term Memory (LSTM) has been proven to be effective in modeling sequential data~\cite{shu2017fake}. However, Transformers, due to their attentive nature, create a better text representation that includes vital information about the input in an efficient manner~\cite{vaswani2017attention}. Thus, in this work, we use Bidirectional Encoder Representations from Transformers (BERT) in creating the representation vector for the news content.

BERT is a pre-trained language model that uses transformer encoders to perform various NLP-based tasks such as text generation, sentiment analysis, and natural language understanding. Previous research have used BERT to achieve state-of-the-art performance on different applications of sentiment analysis tasks~\cite{vaswani2017attention,xu2019bert,li2020enhancing}. Although BERT has been pre-trained on a large corpus of textual data, it should be fine-tuned to perform well on a specific task~\cite{xu2019bert,liu2019roberta}. In this paper, to create a well-defined news content representation, we fine-tune BERT using news articles $\mathbf{x} = \{ w_1, w_2, ..., w_K \}$ from source domain dataset $\mathcal{D}_s$ and a portion $\gamma$ of the target dataset $\mathcal{D}_t$.

In addition to the news content $\mathbf{x}_i$, we also consider the article's comments and user-news interactions. In the experiments, we show that using auxiliary information leads to better detection performance as the model accounts for user's reliability and feedback on the news article.
Nonetheless, previous studies also have shown that comments on news articles can improve fake news detection~\cite{shu2019defend} by extracting semantic clues confirming or disapproving the authenticity of the content. Due to the fact that not all comments are useful for fake news classification, we use Hierarchical Attention Network (HAN) to encode the comments of a news article. The hierarchical structure of HAN facilitates the importance of every comment, as well as the salient word features. To create a representation for the news article comments, we pre-train HAN by stacking it with a feed-forward neural network classifier. After pre-training HAN, we remove the feed-forward classifier and only use the HAN to encode the news article comments. Note that in case a news article does not have any comments, we use vectors with zero values for comments representation.

Moreover, in addition to the news article comments, we also consider the user-news interactions to improve fake news detection. For a news article $\mathbf{x}$, the user-news interactions $\mathbf{u}$ is a binary vector where $\mathbf{u}_i$ indicates user $i$ has tweeted, re-tweeted, or commented on a tweet about that news. Thus, the user-news interactions vector is a representation of the user behaviour toward a news article. To encode this information, we use a feed-forward neural network that takes the binary vector of user-news interaction as input and returns a representation containing important information about that interactions.

After constructing the representation networks (i.e., HAN, BERT, and the feed-forward neural network) and pre-training BERT and HAN, we concatenate the output of these three components into one vector $E = (h'_{comments} || h'_{content} || h'_{interactions})$ ($||$ indicates concatenation) and pass it to another feed-forward network to combine these information into a single vector $\mathbf{E'}$~\cite{karami2020let}. Once we stacked the representation network with a feed-forward neural network classifier, we train a fake news classifier using the source domain data $\mathcal{D}_s$ and a portion $\gamma$ of the target domain data $\mathcal{D}_t$. After training the fake news classifier $F$, we freeze the weights of the representation network and train a domain classifier $D$ using the representation $\mathbf{E'}$ of the news articles. By the end of this process, we have created three sub-components: (1)~a representation network that encodes news content, comments, and user-news interactions, (2)~a single-domain fake news classifier, and (3)~a domain classifier. In the following section, we discuss the second component of the model (i.e., Figure~\ref{fig:proposed}-RL Setting) that uses reinforcement learning to create a domain adaptive representation using $\mathbf{E'}$.

\subsection{Reinforced Domain Adaptation}
Inspired by the success of RL~\cite{kaelbling1996reinforcement,van2016deep,mosallanezhad2019deep}, we model this problem using RL to automatically learn how to convert textual representations from both source domain data $\mathcal{D}_s$ and target domain data $\mathcal{D}_t$. 
Instead of applying a commonly-used approach of adversarial learning to train both $F$ and $D$ classifiers, we utilized an RL-based technique. RL would transform the representation to a new one such that it works well on the fake news classifier $F$, but not on the $D$ classifier (i.e., the adversary). 
In this approach, the agent interacts with the environment by choosing an action $a_t$ in response to a given state $s_t$. The performed action results in state $s_{t+1}$ with reward $r_{t+1}$. The tuple $(s_t, a_t, s_{t+1}, r_{t+1})$ is called an experience which will be used to update the parameters of the RL agent.

In our RL model, an agent is trained to change the news article representation $\mathbf{E'}$ into a new representation that deceives the domain classifier, but preserves the accuracy for the fake news classifier. The RL agent learns this transition by changing the values in the input vector $\mathbf{E'}$ and receiving feedback from both fake news classifier $F$ and domain classifier $D$. To create an RL setting, we define four main parts of RL in our problem, i.e., environment, state, action, and reward.

\begin{itemize}
    \item \textbf{Environment}: the environment in our model includes the RL agent, pre-trained fake news classifier $F$, and the pre-trained domain classifier $D$. In each turn, the RL agent performs an action on the news article representation $E'$ (known as state $s_t$) by changing one of its values (performing action $a_t$). The modified representation is passed through classifiers $F$ and $D$ to get their confidence scores to calculate the reward value. Finally, the reward value and the modified representation is passed to the agent as reward $r_{t+1}$ and state $s_{t+1}$, respectively.
    
    \item \textbf{State}: the state is the current news article representation $\mathbf{E'}$ that describes the current situation for the RL agent.
    
    \item \textbf{Actions}: actions are defined as selecting one of the values in news article representation $\mathbf{E'}$ and changing it by adding or subtracting a small value $\sigma$. The total number of actions are $|E'| \times 2$.
    
    \item \textbf{Reward}: since the aim is to nudge the agent into creating a domain adaptive news article representation, the reward function looks into how much the agent was successful in removing the domain-specific features from the input representation. Specifically, we define the reward function at state $s_{t+1}$ according to the confidence of both the domain classifier and the fake news classifier. Considering the news article embedding $\mathbf{E'}_{t+1}$ with its label $i$ (being fake or real) and domain label $j$ at state $s_{t+1}$, we formally define the reward function as follows:
    \begin{equation}
        r_{t+1}(s_{t+1}) = \alpha Pr_F(l=i|\mathbf{E'}_{t+1}) - \beta Pr_D(l=j|\mathbf{E'}_{t+1})
        \label{eq:reward}
    \end{equation}
    where $l$ indicates the label.
\end{itemize}

\subsection{Optimization Algorithm}

Given the RL setting, the aim is to learn the optimal action selection strategy $\pi (s_t, a_t)$. Algorithm~\ref{alg:opt}\footnote{The source code will become publicly available upon acceptance.} shows this optimization process. At each timestep $t$, the RL agent changes one of the values in the news article representation, $s_t = \mathbf{E'}_t$, and gets the reward value, $r_{t+1}$, based on the modified representation $s_{t+1}=\mathbf{E'}_{t+1}$ and the selected action $a_t$. The goal of the agent is to maximize its reward according to \autoref{eq:reward}.

\begin{algorithm}[t]
	\caption{The Learning Process of \textsc{\m}}\label{learning}
	\begin{algorithmic}[1]
		\REQUIRE~~ News article representations $\mathbf{E' \in \mathcal{D}}$ - fake news classifier $F$ - domain classifier $D$ - parameters $\alpha,\beta$, and $\lambda$ - terminal time $T$.
        \STATE Initialize state $s_t$ and memory $M$.
        \WHILE{training is not terminal}
            \STATE $s_t \gets \mathbf{E'}$
            \FOR{$t \in \{0, 1, ..., T\}$}
                \STATE Choose action $a_t$ according to current distribution $\pi(s_t)$
                \STATE Perform $a_t$ on $s_t$ and get $(s_{t+1}, r_{t+1})$
                \STATE $M \gets M + (s_t, a_t, r_{t+1}, s_{t+1})$
                \STATE $s_{t} \gets s_{t+1}$
                \FOR{each timestep $t$, reward $r$ in $M_t$}
                \STATE $G_t \gets \sum_{i=1}^t \lambda^i r_{i+1}$ 
                \ENDFOR
                \STATE Calculate policy loss according to \autoref{eq:policy_loss}
                \STATE Update the agent's policy according to \autoref{eq:policy_grad}
            \ENDFOR
        \ENDWHILE
	\end{algorithmic}
	\label{alg:opt}
\end{algorithm}

To train the agent, we use the REINFORCE algorithm which uses policy gradient to update the agent~\cite{zhang2020sample}. Considering the agent's policy according to parameters $\theta$ as $\pi_{\theta} (s_t, a_t)$, the REINFORCE algorithm uses the following loss function to evaluate the agent:

\begin{equation}
    \mathcal{L}(\theta) = \log (\pi_\theta(s_t, a_t) \cdot G_t)
    \label{eq:policy_loss}
\end{equation}
where $G_t=\sum_{i=1}^t \lambda^i r_{i+1}$ is the cumulative sum of discounted reward, and $\lambda$ indicates the discount rate. The gradient of the loss function is used to update the agent:

\begin{equation}
    \nabla \theta = \text{lr} \nabla_\theta  \mathcal{L}(\theta)
    \label{eq:policy_grad}
\end{equation}
where $lr$ indicates the learning rate.

\section{Experiments}
In the designed experiments, we try to understand how the differences between domains affect the performance of fake news detection models. Moreover, we investigate how the proposed RL-based approach will overcome performance degradation due to the differences in news domains. Our main evaluation questions are as follows:

\begin{itemize}
    \item \textbf{Q1.} How well does the current fake news detection methods perform on social media data?
    
    \item \textbf{Q2.} How well does the proposed model detect fake news on a \textit{target domain $\mathcal{D}_t$} after training the model on a \textit{source domain $\mathcal{D}_s$}?
    
    \item \textbf{Q3.} How do auxiliary information contribute to the improvement of the fake news detection performance?
\end{itemize}

\textbf{Q1}~evaluates the quality of fake news detection models. We answer this question by training and testing fake news detection models on the same domain and comparing their performance. \textbf{Q2}~studies the effect of domain differences on the performance of fake news detection models. We use a similar approach to \textbf{Q1} to answer this question by training on one domain, but testing on another. In \textbf{Q3}~we perform ablation studies to analyze the impact of auxiliary information, the reward function parameters $\alpha$ and $\beta$, and the portion of target domain data $\gamma$ needed to achieve an acceptable detection performance.

\subsection{Datasets}
We use the well-known fake news data repository FakeNewsNet~\cite{shu2020fakenewsnet} which contains news articles along with their auxiliary information such as users' metadata and news comments. These news articles have been fact-checked with two popular fact-checking platforms -  \textit{Politifact} and \textit{GossipCop}. Politifact fact-checks news related to the U.S. political system, while GossipCop fact-checks news from the entertainment industry. In addition to the existing Politifact news articles from FakeNewsNet, we enrich the dataset by adding $5,000$ annotated Politifact news from the dataset introduced by Rashkin et al. \cite{rashkin2017truth}. \autoref{tab:data_stats} shows the statistics of the final dataset. The Politifact news articles from \cite{rashkin2017truth} includes truth rating from $0$ to $5$, in which we only consider news with $\text{label} \in \{0, 4, 5\}$.

\begin{table}[t]
\centering
\begin{tabular}{l|cc}
\hline\hline
                      & \textbf{Politifact}    & \textbf{GossipCop}   \\ \hline
\# True News          & 2,645         & 3,586       \\
\# Fake News          & 2,770         & 2,230       \\
\# News               & 5,415         & 5,819       \\
\# News with Comments & 415           & 5,819       \\
\# Users              & 60,053        & 43,918      \\
\# Unique Users       & \multicolumn{2}{c}{100,520} \\\hline\hline
\end{tabular}
\caption{The statistics of \textit{Politifact} and \textit{GossipCop} datasets. The \textit{Politifact} dataset contains news articles from both FakeNewsNet and the dataset provided by Rashkin et al.~\cite{rashkin2017truth}, while \textit{GossipCop} contains news articles from FakeNewsNet only.}
\label{tab:data_stats}
\end{table}

\subsection{Data Pre-processing}
Each news articles from the final dataset contains news content, users' comments, and their meta-data. We pre-process the textual data (i.e., news content and users' comments) by removing the punctuation, mentions, and out-of-vocabulary words. Further, as BERT has a limitation of getting $512$ words as input, we truncate the news content and every comment to include its first $512$ words. Finally, we utilize the users' meta-data to create user-news interaction matrix by tracking every user's interactions with news articles.

\subsection{Implementation Details}
The training process has been conducted in three stages: (1)~pre-training the representation networks, (2)~training the fake news and domain classifiers, and (3)~training the RL agent. In what follows, we will expand the implementation details of each stage.

\noindent \textbf{Pre-training:} In this stage, we fine-tune BERT and HAN networks to generate a reasonable text representation from news content and users' comments, respectively. Motivated by the low memory consumption of the distilled version of Bert~\cite{sanh2019distilbert}, we used the base model of Distilled BERT for creating the textual representation of the news contents. The model was fine-tuned for $3$ iterations using a classifier on top of it. Moreover, to create the representations related to the user's comments, we pre-trained HAN by stacking it with a simple fake news classifier and fine-tuning it for $5$ iterations. After fine-tuning both models, the classifier module in both distilled Bert and HAN were removed. These pre-trained networks were placed in the final architecture of our model.

\noindent \textbf{Training classifiers $F$ and $D$:} With passing the news content through the representation network in \autoref{fig:proposed}, we trained a fake news classifier and a domain classifier using the news content representations. During the training, we did not update the weights of BERT and HAN networks. In this stage, a dropout with $p=0.2$ was used for both fake news and domain classifiers. Both classifiers use a similar feed-forward neural network with a single hidden layer of $256$ neurons. The networks are trained using Adam optimizer with a learning rate of $1e-5$ and a Cross Entropy loss function:
\begin{equation}
    \mathcal{L}_{CE} = -\frac{1}{M} \sum_{i=1}^{M} (y_i \log (p_i) + (1-y_i) \log (1-p_i))
    \label{eq:ce_loss}
\end{equation}

\begin{table*}[ht]
\centering
\small
\begin{tabular}{l|cc|cc|cc|cc} 
\hline \hline
\multirow{2}{*}{\textbf{Model}} & \multicolumn{2}{c|}{\textbf{GossipCop $\rightarrow$ Politifact}}          & \multicolumn{2}{c|}{\textbf{GossipCop $\rightarrow$ GossipCop}}          & \multicolumn{2}{c|}{\textbf{Politifact $\rightarrow$ Politifact}}          & \multicolumn{2}{c}{\textbf{Politifact $\rightarrow$ GossipCop}}            \\
                       & \textbf{F1}                                    & \textbf{AUC}                      & \textbf{F1}                                    & \textbf{AUC}                      & \textbf{F1}                                    & \textbf{AUC}                      & \textbf{F1}                                    & \textbf{AUC}                       \\ 
\cline{1-9}
CSI                    & 0.532  $\pm$  0.11   & 0.563 $\pm$ 0.14         & 0.811~$\pm$~0.05         & 0.832 $\pm$ 0.08         & 0.756$\pm$0.02         & 0.785 $\pm$ 0.10         & 0.589$\pm$0.10         & \uline{0.610} $\pm$ 0.13  \\
URG                    & 0.448$\pm$0.11         & 0.450 $\pm$ 0.11         & 0.770$\pm$0.04         & 0.792 $\pm$ 0.10         & 0.526$\pm$0.01         & 0.741 $\pm$ 0.08         & 0.460$\pm$0.09         & 0.532 $\pm$ 0.15          \\
DEF                    & 0.621$\pm$0.10         & 0.687 $\pm$ 0.11         & 0.857$\pm$0.03         & 0.893 $\pm$ 0.07         & 0.768$\pm$0.01         & 0.793 $\pm$ 0.06         & 0.423$\pm$0.06         & 0.561 $\pm$ 0.11          \\
BBL                    & 0.695$\pm$0.07 & 0.715 $\pm$ 0.09 & \textbf{0.918}$\pm$0.02         & \uline{0.947} $\pm$ 0.08         & 0.824$\pm$0.02 & \textbf{0.868} $\pm$ 0.04         & 0.558$\pm$0.09         & 0.597 $\pm$ 0.05          \\
UDA                    & 0.673$\pm$0.08         & 0.695 $\pm$ 0.09         & 0.901$\pm$0.02         & 0.923 $\pm$ 0.06         & 0.727$\pm$0.02         & 0.762 $\pm$ 0.02         & \uline{0.596}$\pm$0.09 & 0.602 $\pm$ 0.07          \\
EMB                    & \uline{0.704}$\pm$0.07              & \uline{0.719} $\pm$ 0.08              & \uline{0.914}$\pm$0.04              & \textbf{0.952} $\pm$ 0.05              & \uline{0.835}$\pm$0.04              & \uline{0.852} $\pm$ 0.03              & 0.586$\pm$0.06              & 0.601 $\pm$ 0.05               \\
SRE                    & 0.656$\pm$0.09         & 0.714 $\pm$ 0.11         & 0.867$\pm$0.05         & 0.931 $\pm$ 0.07         & 0.720$\pm$0.01         & 0.751 $\pm$ 0.05         & 0.429$\pm$0.08         & 0.521 $\pm$ 0.09          \\
\m               & \textbf{0.726}$\pm$0.11         & \textbf{0.728} $\pm$ 0.10         & 0.905$\pm$0.10 & 0.933 $\pm$ 0.08 & \textbf{0.844}$\pm$0.05         & 0.810 $\pm$ 0.08 & \textbf{0.601}$\pm$0.09         & \textbf{0.612} $\pm$ 0.11          \\
\hline \hline
\end{tabular}
\caption{\textit{Single-domain:} Performance of \m and baselines, trained \textit{only on one domain} and tested on both \textit{Politifact} and \textit{GossipCop} domains (Train $\rightarrow$ Test). The values indicate \textit{F1} and \textit{AUC} measures on the test set.}
\label{table:no_cross_F1_AUC}
\end{table*}

\begin{figure*}[ht!] \vspace{-10pt}
\begin{subfigure}{.48\textwidth}
  \centering
  \includegraphics[width=.68\linewidth]{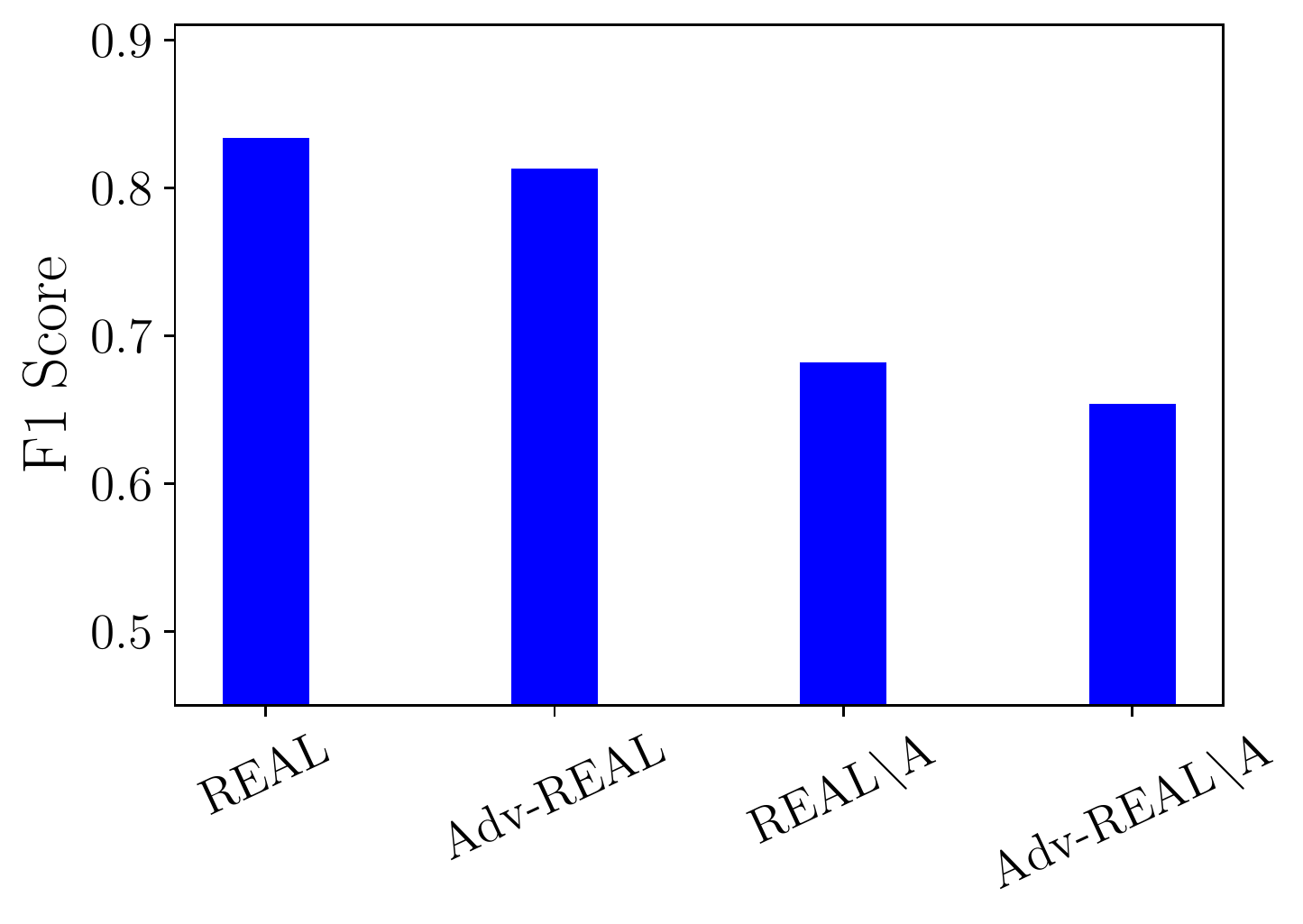}  
  \caption{GossipCop as source, Politifact as target domain}
  \label{fig:politifact_target}
\end{subfigure}
\begin{subfigure}{.48\textwidth}
  \centering
  \includegraphics[width=.68\linewidth]{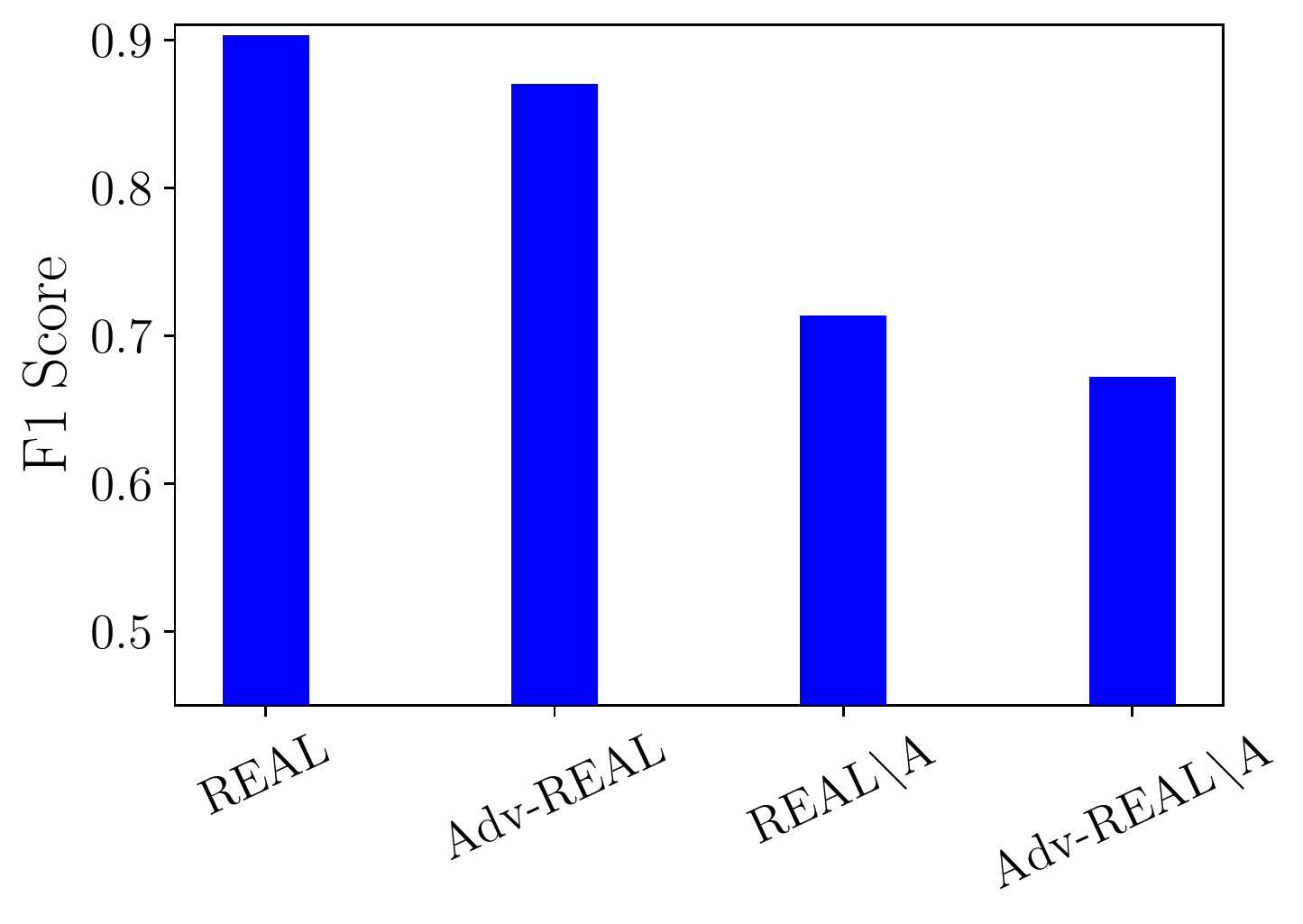}  
  \caption{Politifact as source, GossipCop as target domain}
  \label{fig:gossipcop_target}
\end{subfigure}
\caption{The impact of auxiliary information and reinforcement learning for cross-domain fake news detection. Different versions of \m are evaluated on the target domain.}
\label{fig:components}
\end{figure*}

\noindent \textbf{Training the RL agent:} Once finished with the first two stages, the fake news classifier and the domain classifier are placed as the reward function of the RL setting to train the agent. We train the agent using Algorithm~\ref{alg:opt} for $2,000$ episodes performing only $T=20$ actions. As the long term reward function is important, we used a discount rate of $\lambda=0.99$. In updating the agent network, we applied Adam optimizer and set the parameters with $\alpha=\beta=0.5$, $\sigma=0.01$, and learning rate $lr=1e-4$. The RL agent uses a feed-forward neural network with $2$ hidden layers of $512$ and $256$ neurons.

\subsection{Baselines}
For evaluating the effectiveness of \m on fake news detection task, we compare it to the baseline models described below. For a comprehensive comparison we consider state-of-the-art baselines that (1)~only use news content (BERT-BiLSTM and BERT-UDA), (2)~use both news content and users' comments (TCNN-URG and dEFEND), (3)~consider users' interactions (CSI), and (4)~considers propagation network (RoBERT-EMB). In addition to these baselines, we consider two variants of \m, Simple-\m and Adv-\m that use bi-directional GRU instead of BERT/HAN and adversarial training, respectively. These two baselines help us to study the effectiveness of using a complex architecture such as BERT and RL for domain adaptive fake news detection.

\begin{itemize}
    \item \textbf{TCNN-URG (URG)}~\cite{qian2018neural}: Based on TCNN~\cite{kim2014convolutional}, this model uses convolutional neural networks to capture various granularity of news content as well as including users' comments. 
    
    \item \textbf{CSI}~\cite{ruchansky2017csi}: CSI is a hybrid model that utilizes news content, users' comments, and the news source. The news representation is modeled using LSTM neural network using Doc2Vec~\cite{le2014distributed} that outputs an embedding for both news content and users' comments. For fair comparison, we disregarded the news source feature.
    
    \item \textbf{dEFEND (DEF)}~\cite{shu2019defend}: This model uses a co-attention network to model both news content and users' comments. dEFEND captures explainable content information for fake news detection.
    
    \item \textbf{BERT-UDA (UDA)}~\cite{gong2020unified}: This model uses feature-based and instance-based domain adaptation on BERT to create domain-independent news content representation.
    
    \item \textbf{BERT-BiLSTM (BBL)}~\cite{vlad2019sentence}: This model uses a complex neural network model including BERT~\cite{vaswani2017attention}, bi-directional LSTM layer, and a capsule layer to classify news content. This model uses transfer learning to work across different domains.
    
    \item \textbf{RoBERTa-EMB (EMB)}~\cite{silva2021embracing}: This model adopt instance-based domain adaptation. It uses RoBERTa-base~\cite{liu2019roberta} to create news content representation, while uses an unsupervised neural network to create the propagation network's representation~\cite{silva2020embedding}. To evaluate this method on single-domain, we disable the domain adaptation process.
    
    \item \textbf{Simple-\m (SRE)}: To study the effect of using complex networks such as BERT and HAN for creating the representation of the news content and users comments, we replace BERT and HAN with a bi-directional Gated Recurrent Unit (BiGRU) stacked with a location-based attention layer~\cite{chung2014empirical}. Specifically, we use two separate BiGRU neural networks to create $h'_{content}$ and $h'_{comments}$. To create $h'_{comments}$, we concatenate all of the comments and select the first $256$ words.
    
\end{itemize}

\subsection{Experimental Results}
\begin{figure*}[ht]\vspace{-10pt}
\begin{subfigure}{.32\textwidth}
  \centering
  \includegraphics[width=0.85\linewidth]{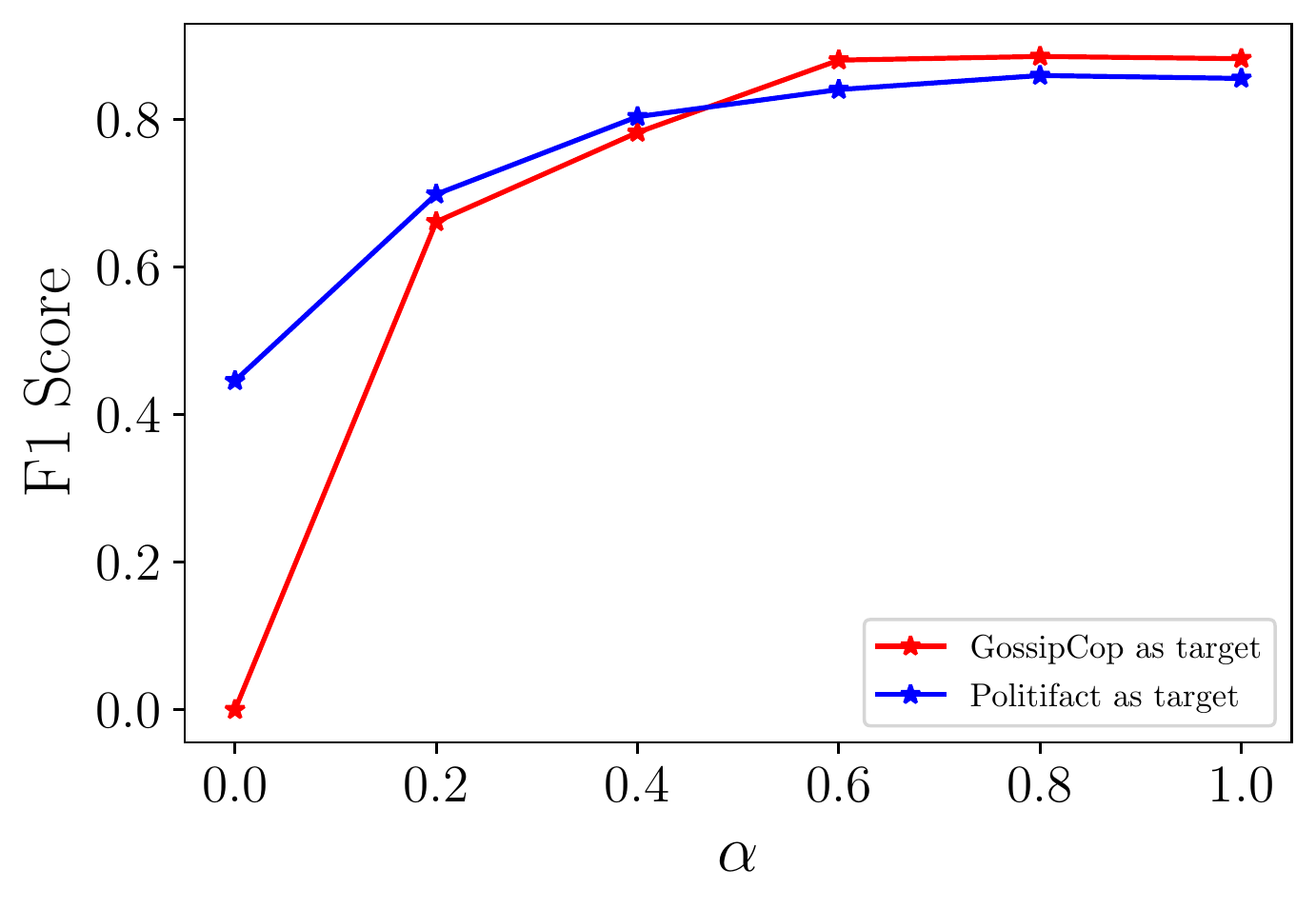}  
  \caption{Effect of the $\alpha$ parameter}
  \label{fig:alpha}
\end{subfigure}
\begin{subfigure}{.32\textwidth}
  \centering
  \includegraphics[width=0.85\linewidth]{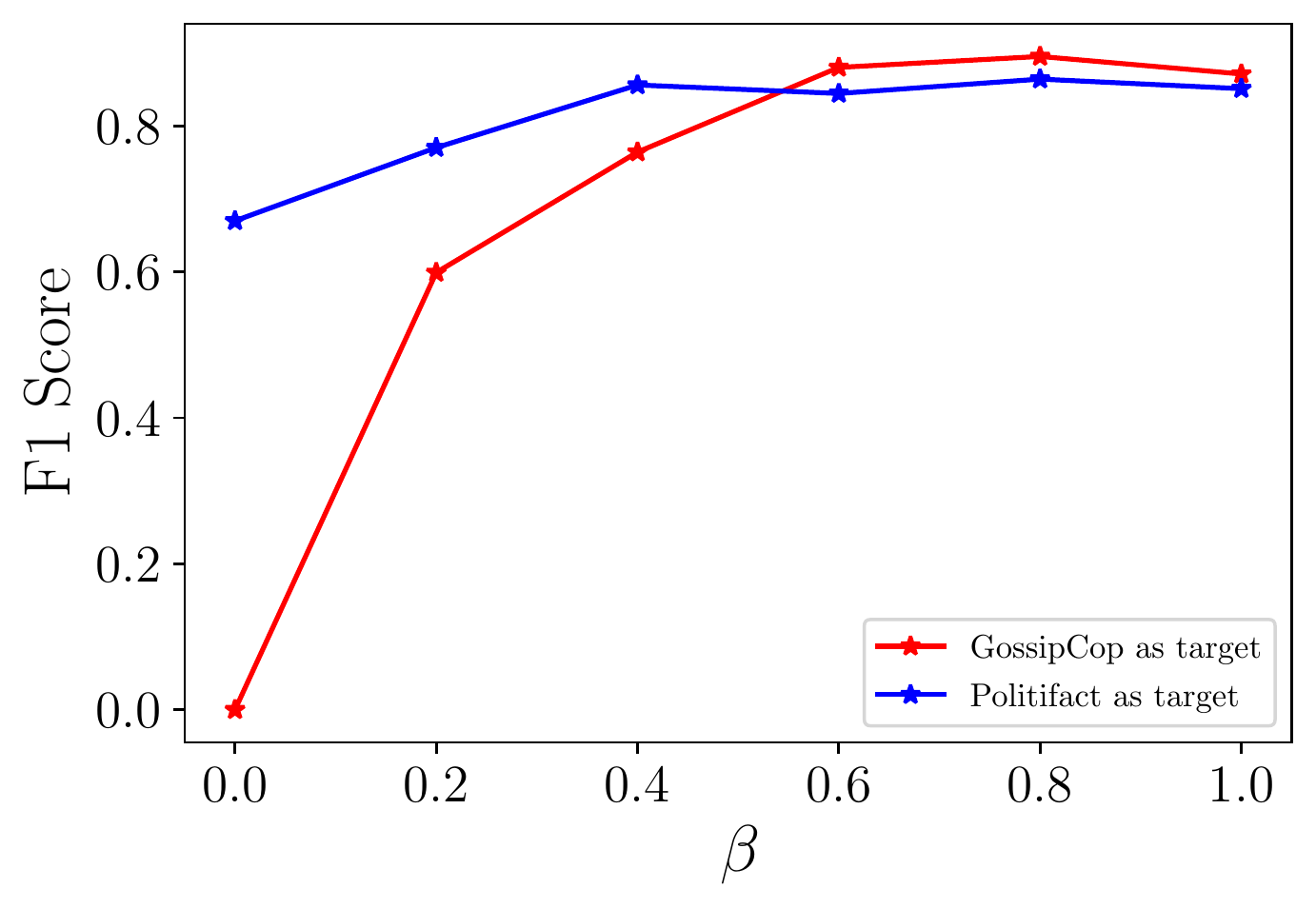}  
  \caption{Effect of the $\beta$ parameter}
  \label{fig:beta}
\end{subfigure}
\begin{subfigure}{.32\textwidth}
  \centering
  \includegraphics[width=0.85\linewidth]{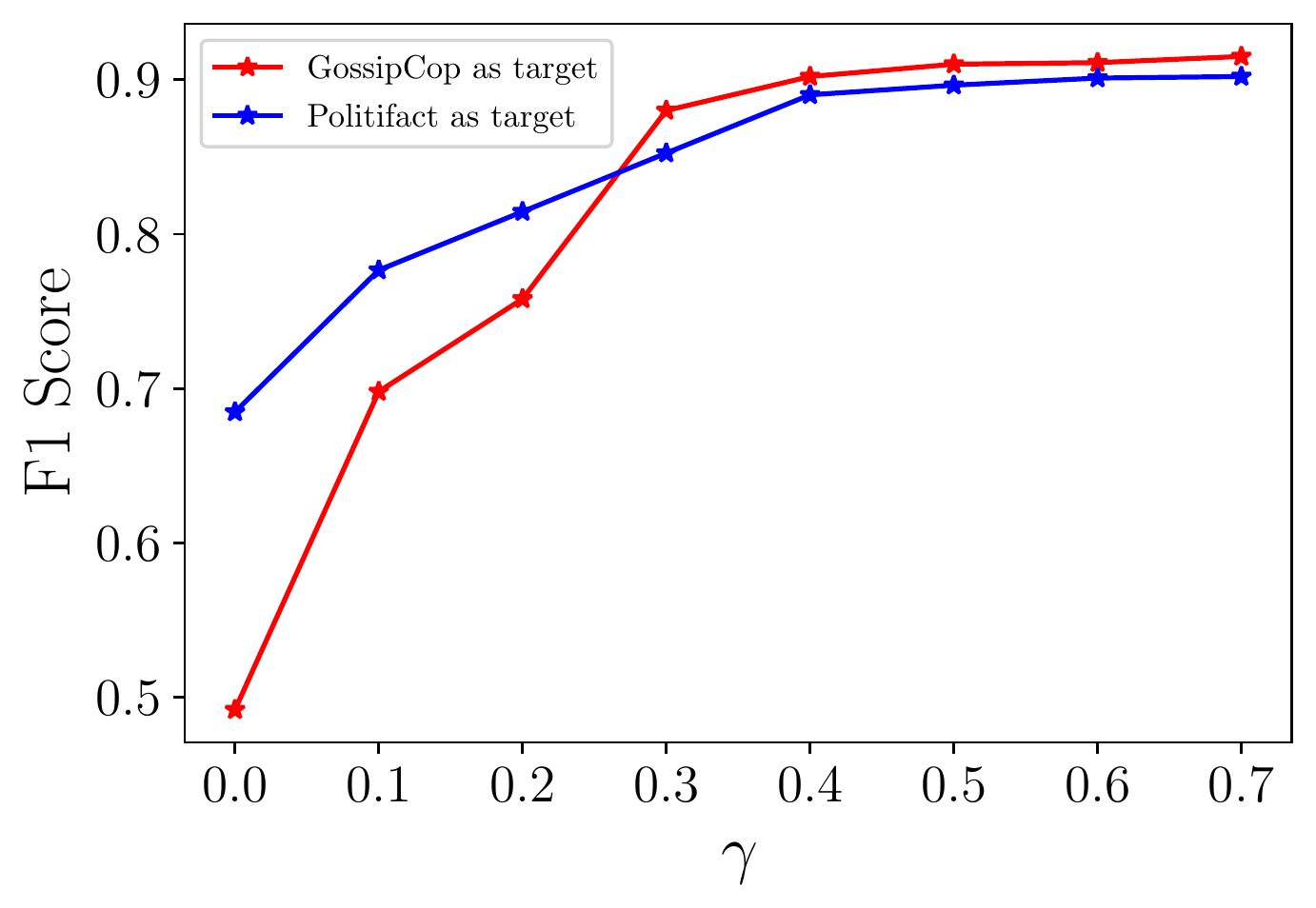}  
  \caption{Effect of the target domain portion $\gamma$}
  \label{fig:gamma}
\end{subfigure}
\caption{The impact of reward function parameters and the portion of the target domain data. Blue line shows the result of \m on \textit{GossipCop}, while the red line shows the F1 for \m trained on the \textit{Politifact} domain.}
\label{fig:params}
\end{figure*}

To answer questions \textbf{Q1} and \textbf{Q2}, we trained \m and the baselines on a source domain $S$ and tested them on both source domain $S$ and target domain $T$. We used $k$-fold validation and calculated the average AUC and F1 scores. For single-domain and cross-domain, we set the number for folds as $k= 9$ and $k= 10$, respectively. In single-domain training, we use the source domain data $\mathcal{D}_s$, while for training \m and the baselines in a cross-domain setting, we use the source domain data $\mathcal{D}_s$ combined with portion $\gamma$ of the target domain data $\mathcal{D}_t$. In the subsequent subsection it will be shown in \autoref{fig:gamma} that performance improvements taper off after using $30\% (\gamma=0.3)$ of the target domain data. Finally, to answer \textbf{Q3}, we perform an ablation study to measure the impact of using auxiliary information and reinforcement learning.

\noindent \textbf{Fake News Detection Results (Q1).}
\autoref{table:no_cross_F1_AUC} shows a comparison of the baselines with \m. Training is applied on a single domain dataset $\mathcal{D}_s$ and tested on both source and target domains. For this experiment, we removed the domain classifier's feedback from the RL agent by setting the parameter $\beta=0$. From the results we conclude that (1)~all baseline models and \m perform reliably well when both training and testing news come from a single domain, and (2)~due to the considerable decrease in performance when testing with news from another domain, it appears that the \textit{Politifact} and \textit{GossipCop} news domains have different properties. These results imply that the evaluated models are not agnostic to domain differences. \m performed better than baselines for the majority of scenarios. The only case where \m is under-performed is when the model is trained and tested on the \textit{GossipCop} dataset. The better performance of \m on the \textit{Politifact} domain suggests that BBL may have overfitting issues and \m benefits the use of auxiliary information for creating a more general fake news classifier that performs well on both domains. It is worth mentioning that EMB and BBL models are similar to each other in terms of model architecture except that EMB utilizes the user's propagation network as well.
Comparing the results between these two models also reveals that using auxiliary information can be helpful in detecting fake news.

\noindent \textbf{Cross-Domain Results (Q2).} 
\autoref{tab:cross_domain_results} shows the performance of models on the target domain. In this experiment, we used $\gamma=0.3$ of nine-folds of the target domain dataset $\mathcal{D}_t$ in addition to all the source domain dataset $\mathcal{D}_s$. Performance measures are calculated based on the tenth-fold of the target domain dataset.
The results indicate that \m detects fake news in the target domain most efficiently, in comparison to the baselines. For example, compared with the best baselines, both F1 and AUC scores have improved. This indicates that most baselines suffer from overfitting on the source domain. Moreover, the results show that using a small portion of the target domain dataset can lead to a large increase in the cross-domain classification, indicating that the RL agent learns more domain-independent features by the feedback received from the domain classifier. It is also notable that Simple-\m (SRE) perform surprisingly well despite using weaker network architecture than the most baselines. 

\noindent \textbf{Impact of RL and Auxiliary Information (Q3).} To show the relative impact using reinforcement learning and auxiliary information, we created the following variants of \m:

\begin{itemize}
    \item[-] \textbf{\m$\backslash$A}: To study the effects of using auxiliary information in fake news detection, we created a variant of \m which does not use users' comments and user-news interactions. Thus, this model only uses BERT to create news article representations.
    
            \begin{table}[]
\centering
\small
\begin{tabular}{l|cc|cc}
    \hline\hline
    \multirow{2}{*}{{\bf Model}} & \multicolumn{2}{c|}{{\bf GossipCop $\rightarrow$ {\bf Politifact}}} & \multicolumn{2}{c}{{\bf Politifact $\rightarrow$ {\bf GossipCop}}} \\
    & \textbf{AUC} & \textbf{F1} & \textbf{AUC} & \textbf{F1} \\
    \hline
    
    CSI & 0.581 $\pm$ 0.03 & 0.547 $\pm$ 0.02 & 0.612 $\pm$ 0.03 & 0.598 $\pm$ 0.02 \\
    URG & 0.503 $\pm$ 0.04 & 0.486 $\pm$ 0.02 & 0.545 $\pm$ 0.04 & 0.471 $\pm$ 0.02 \\
    DEF & 0.712 $\pm$ 0.02 & 0.634 $\pm$ 0.01 & 0.583 $\pm$ 0.02 & 0.583 $\pm$ 0.01 \\
    BBL & 0.748 $\pm$ 0.01 & 0.711 $\pm$ 0.01 & 0.634 $\pm$ 0.01 & 0.634 $\pm$ 0.01 \\
    UDA & 0.812 $\pm$ 0.02 & 0.778 $\pm$ 0.01 & 0.702 $\pm$ 0.01 & 0.702 $\pm$ 0.01 \\
    EMB & 0.870 $\pm$ 0.01 & 0.876 $\pm$ 0.02 & \underline{0.846} $\pm$ 0.03 & \underline{0.795} $\pm$ 0.03 \\
    SRE & \underline{0.885} $\pm$ 0.03 & \underline{0.881} $\pm$ 0.04 & 0.838 $\pm$ 0.03 & 0.791 $\pm$ 0.05 \\
    \m & {\bf 0.901} $\pm$ 0.02 & {\bf 0.892} $\pm$ 0.03 & {\bf 0.862} $\pm$ 0.05 & {\bf 0.815} $\pm$ 0.04 \\
\hline\hline
\end{tabular}
\caption{\label{tab:cross_domain_results} \textit{Cross-domain:} Cross-domain fake news detection results on the target domain. \textbf{source} $\rightarrow$ \textbf{target} indicates the models have been trained on the \textbf{source} domain and tested on the \textbf{target} domain. In this case, we use the source domain data in addition to a small portion of the target domain data.}
\end{table}

    \item[-] \textbf{Adv-\m (ARE)}: To study the effect of the RL agent in domain adaptation, we use adversarial training to create a domain adaptive news article representation. In this model, we use the representation network, fake news classifier $F$, ARE domain classifier $D$ in an adversarial setting using the following loss function to train a fake news classifier and representation network that is capable of creating domain independent features.
    \begin{equation}
    \label{eq:adv_loss}
        \min_F \max_D \mathcal{L}_{CE}(F) - \mathcal{L}_{CE}(D),
    \end{equation}
    
    where $L$ is the binary cross entropy loss similar to \autoref{eq:ce_loss}.
    In this model we do not need to pre-train the fake news classifier $F$ and the domain classifier $D$. Instead, we train the representation network, fake news classifier $F$, and domain classifier $D$ as a whole using \autoref{eq:adv_loss}.

    \item[-] \textbf{Adv-\m$\backslash$A}: In this model, we removed both RL component and the auxiliary information.
\end{itemize}

According to \autoref{fig:components}, it is notable that removing the users' comments, and the user-news interactions severely impacts the performance of cross-domain fake news detection. Although using adversarial training for cross-domain fake news detection performs well, reinforcement learning allows us to adopt the news article representation using any pre-trained domain and fake news classifier in the reward function without considering its differentiability.  

\subsection{Parameter Analysis}
\autoref{fig:params} shows the results of varying parameters in the reward function (i.e., \autoref{eq:reward}) and the portion of target domain data. We perform sensitivity analysis on the reward function parameters $\alpha, \beta$ by fine-tuning the parameters across the range $[0.0, 1.0]$. We also examine the effect of the $\gamma$ parameter by changing it across the range $[0.0,0.7]$. We evaluate the effect of varying these parameters by evaluating the F1 score of cross-domain fake news detection. 

\autoref{fig:alpha} shows the F1 score for different values of $\alpha$ when $\beta = 0.5$. The $\alpha$ parameter controls the amount of the contribution of the fake news classifier in the reward function. This figure indicates that the performance does not increase when $\alpha \geq 0.6$ and even lead to a decrease in the performance if we use a higher value than $\beta$. 

\autoref{fig:beta} illustrates the results of varying $\beta$ when $\alpha=0.5$. By using $\beta=0.0$, \m does not consider the feedback of the domain classifier, thus, the performance on the target domain will be dismal. By increasing $\beta$, RL penalizes the agent more when the agent cannot decrease the confidence of the domain classifier. Similar to the $\alpha$ parameter, the performance does not increase by a large margin by using $\beta \ge 0.6$.

To analyze the impact of $\gamma$, we use various portions of the target domain dataset to train \m. Intuitively, using all the data from target domain will improve the performance; however, in real-world scenarios, we may not have access to a comprehensive dataset from the target domain and using such dataset can lead to an expensive computation time. In real-world scenarios, domain-specific data is usually sparse and training on $100\%$ of the target domain's data is impractical and unrealistic. Based on our analysis, using $30\%$ of target domain dataset seems to be sufficient for training a reliable domain adaptive fake news detection model.

\section{Conclusion and Future Work}

Collecting and integrating news articles from different domains and providing human annotations by fact-checking the contents for the purpose of aggregating a dataset are resource-intensive activities that hinder the effective training of automated fake news detection models. Although many deep learning models have been proposed for fake news detection and some have exhibited good results on the domain they were trained on, we show in this work that they have limited effectiveness in other domains. To overcome these challenges, we propose the \textbf{RE}inforced domain \textbf{A}daptation \textbf{L}earning for \textbf{F}ake \textbf{N}ews \textbf{D}etection (\m) task which could effectively classify fake news on two separate domains using only a small portion of the target domain data. Further, \m also leverages auxiliary information to enhance fake news detection performance. Experiments on real-world datasets show that in comparison to the current SOTA, \m adopts better to a new domain by using auxiliary information and reinforcement learning and achieves high performance in a single-domain setting.  

Enhancements on \m could focus on cross-domain fake news detection under limited supervision to address the effects of limited or imprecise data annotations by applying weak supervision learning in a domain adaptation setting. Additionally, \m could benefit from methods that automatically defend against adversarial attacks. For example, malicious comments that promote a fake news article could degrade the performance of \m because of \m's contingency on users' comments as auxiliary information. 
Finally, we also propose addressing the inconsistencies in multi-modal information due to the breadth of the types and sources of information that are required for a cross-domain fake news classifier such as \m. 



\section{Acknowledgements}
This material is based upon work supported by ONR (N00014-21-1-4002), and the U.S. Department of Homeland Security under Grant Award Number 17STQAC00001-05-00\footnote{Disclaimer: “The views and conclusions contained in this document are those of the authors and should not be interpreted as necessarily representing the official policies, either expressed or implied, of the U.S. Department of Homeland Security.”
}. Kai Shu is supported by the NSF award \#2109316.


\bibliographystyle{ACM-Reference-Format}
\bibliography{sample-base}

\end{document}